\documentclass[preprint,preprintnumbers,amsmath,amssymb,superscriptaddress]{revtex4}
\usepackage{txfonts}
\usepackage{graphicx}
\usepackage{dcolumn}
\usepackage{bm}
\usepackage{array}
\usepackage{changes}
\usepackage{verbatim}
\usepackage{upgreek}
\usepackage[colorlinks=true,plainpages=false,linkcolor=blue,urlcolor=blue,citecolor=blue,pdfpagemode=UseNone,pdfstartview=FitBH]{hyperref}

\begin{document}
	
	\title{Antiferromagnetic spin fluctuations and unconventional superconductivity in topological superconductor candidate YPtBi revealed by $^{195}$Pt-NMR}
	\author{Y. Z. Zhou}
	\affiliation{Institute of Physics, Chinese Academy of Sciences,\\
		and Beijing National Laboratory for Condensed Matter Physics,Beijing 100190, China}
	\affiliation{School of Physical Sciences, University of Chinese Academy of Sciences, Beijing 100190, China}
	
    \author{J. Chen}
	\affiliation{Institute of Physics, Chinese Academy of Sciences,\\
		and Beijing National Laboratory for Condensed Matter Physics,Beijing 100190, China}
	
    \author{Z. X. Li}
	\affiliation{School of Physical Science and Technology, ShanghaiTech University, Shanghai, 201210, China}

    \author{J. Luo}
	\affiliation{Institute of Physics, Chinese Academy of Sciences,\\
		and Beijing National Laboratory for Condensed Matter Physics,Beijing 100190, China}

    \author{J. Yang}
	\affiliation{Institute of Physics, Chinese Academy of Sciences,\\
		and Beijing National Laboratory for Condensed Matter Physics,Beijing 100190, China}

	\author{Y. F. Guo}
	\affiliation{School of Physical Science and Technology, ShanghaiTech University, Shanghai, 201210, China}
    \affiliation{ShanghaiTech Laboratory for Topological Physics, Shanghai 201210, China}

    \author{W. H. Wang}
	\affiliation{Institute of Physics, Chinese Academy of Sciences,\\
		and Beijing National Laboratory for Condensed Matter Physics,Beijing 100190, China}

	\author{R. Zhou}
	\email{rzhou@iphy.ac.cn}
	\affiliation{Institute of Physics, Chinese Academy of Sciences,\\
		and Beijing National Laboratory for Condensed Matter Physics,Beijing 100190, China}
	\affiliation{Songshan Lake Materials Laboratory, Dongguan, Guangdong 523808, China}

	\author{Guo-qing Zheng}
	\affiliation{Department of Physics, Okayama University, Okayama 700-8530, Japan}

	\date{\today}
	
	\begin{abstract}
		{We report $^{195}$Pt nuclear magnetic resonance (NMR) measurements on topological superconductor candidate YPtBi which has the broken inversion symmetry and topological non-trivial band structures due to the strong spin-orbit coupling(SOC). In the normal state, we find that Knight shift $K$ is  field- and temperature-independent, suggesting that the contribution from the topological bands is very small at low temperatures. However, the spin-lattice relaxation rate 1/$T_1$ divided by temperature ($T$), 1/$T_1T$, increases with decreasing $T$, implying the existence of antiferromagnetic spin fluctuations. In the superconducting state, no Hebel-Slichter coherence peak is seen below $T_{\rm c}$ and 1/$T_1$ follows $T^{3}$ variation, indicating the unconventional superconductivity.  The finite spin susceptibility at zero-temperature limit and the anomalous increase of the NMR line width below $T_{\rm c}$ point to a mixed state of spin-singlet and spin-triplet(or spin-septet) pairing. 
		}
	\end{abstract}

	
	
	\maketitle
	Topological superconductors (TSCs) have application potential in quantum computing, due to the existence of Majorana fermions\cite{PhysRevLett.94.166802,Topologicalinsulators,Topologicalinsulatorsandsuperconductors}. Over the last two decades, only a few TSCs candidates, such as Cu$_{x}$Bi$_{2}$Se$_{3}$, UTe$_2$, and K$_{2}$Cr$_{3}$As$_{3}$, have emerged, with Cu$_x$Bi$_2$Se$_3$ and K$_2$Cr$_3$As$_3$ being of odd parity\cite{Sato_2017,CuxBi2Se3_NMR,Ran2019,Yang2021}. Besides these, non-centrosymmetric superconductors, where inversion symmetry is broken and thus leads to an antisymmetric spin-orbit coupling term (ASOC) that brings about the spin-triplet component\cite{Li2Pt3B}, were also proposed to be possible TSCs.\cite{PhysRevB.76.045302}.
	
	A series of non-centrosymmetric ternary half-Heusler compounds $R$(Pd, Pt)Bi($R$: rare earth) with heavy elements whose zinc-blende type sublattice resembles (Cd, Hg)Te quantum well structure\cite{PhysRevB.76.045302,HgTequantumwells},  have attracted a lot of attention\cite{Half-Heusler,Tunable}. Strong SOC is also present in those half-Heusler compounds which leads to topologically nontrivial $\Gamma_{6}$/$\Gamma_{8}$ band inversion\cite{PhysRevLett.105.096404,PhysRevB.82.125208},  as confirmed by Angle-resolved photoemission spectroscopy(ARPES) experiments\cite{unusual_topological_surface_states,Metallic_surface_electronic_state}. Most remarkably, superconductivity was discovered in $X$PtBi($X$ = Y, Lu, La)\cite{Superconductivity,LaBiPt_SC,LnBiPt_SC}. Among them, YPtBi attracts special attention. First, the hole carrier concentration is only at the order of $10^{18}$$\sim 10^{19}$cm$^{-3}$\cite{Superconductivity,pressure}, which seems to be too small to explain its $T_{\rm c}$ = 0.77 K by a conventional theory\cite{DOS}. Furthermore,	owing to band inversion, the total angular momentum of low-energy electronic states in $\Gamma_{8}$ band near the chemical potential is $j = $ 3/2, which derives from strong SOC between $s$ = 1/2 spin and $l$ = 1 $p$-orbitals\cite{PhysRevB.76.045302,Sato_2017}. Quasiparticles due to $j$ = 3/2 pairing may carry high angular-momentum, as $(3/2)\otimes(3/2) =0\oplus1\oplus2\oplus3$ corresponds to spin singlet, triplet, quintet, and septet, respectively\cite{PhysRevLett.117.075301,Roy_2019,PhysRevB.96.214514,Venderbos_2018}. Based on this, theoretical studies suggested that YPtBi is a possible topological superconductor\cite{Roy_2019}. While the nature
of the pairing still remains controversial, various symmetries have been proposed, such as $p$+$s$ mixed singlet-septet pairing\cite{London,singlet_septet,Majorana_surface_modes,Inflated_nodes}, $d$+$s$ mixed singlet-quintet pairing\cite{Singlet-quintet,Singlet-quintet_2}, $s$-wave quintet pairing\cite{Boettcher_2018}, and $d$+$s$ singlet pairing\cite{d+s}. 

Although many studies already revealed the topologically nontrivial nature of YPtBi, the low $T_{\rm c}$ and  extremely low carrier concentration hinder the research on its superconductivity using probes such as heat capacity\cite{Heat_apacity}.
Until now, very little information about the superconducting state of YPtBi has been obtained, and the pairing symmetry is still a mystery.
The temperature dependence of the upper critical field $H_{\rm c2}(T)$ suggests that YPtBi is not a simple spin-singlet superconductivity\cite{pressure}. London penetration depth does not show any saturation down to $T \sim 0.06 T_{\rm c}$, suggesting the existence of unusual pairing in YPtBi\cite{London}. Although previous $^{209}$Bi-NMR studies revealed the non-trivial topology of the band structure in YPtBi\cite{Nowak2014,zhang2016}, microscopic studies are still lacking to settle down the pairing symmetry in YPtBi.

In this letter, we investigate the superconductivity of YPtBi by $^{195}$Pt-NMR measurements. In the normal state, we find that 1/$T_1T$ increases with decreasing temperature, implying the appearance of the antiferromagnetic spin fluctuations.
Below $T_{\rm c}$, the spin-lattice relaxation rate 1/$T_1$ decreases as $T^{3}$ with no coherence peak, indicating the unconventional superconducting state. The residual spin susceptibility at zero-temperature limit and the detailed analysis of temperature dependent  1/$T_1$ suggest that the Cooper pairs might be the mixed superconducting state of  spin-singlet and spin-triplet(or spin-septet) pairing.
	
YPtBi single crystals were synthesized out of Bi flux as described in Refs.\cite{zhang2016}. The $T_{\rm c}$ of our sample is 0.77 K as measured by the AC susceptibility using an in situ NMR coil\cite{SM}. In order to obtain enough NMR signal intensity at a very small field in the superconducting state, the single crystals are crushed into fine powders to gain the surface area. Below $T$ = 1.5 K, measurements were conducted by using a $^{3}$He/ $^{4}$He dilution refrigerator. We used a commercial BeCu/NiCrAl clamp cell from C$\&$T Factory Co., Ltd.\cite{Yokogawa2007}. The applied pressure has been calibrated by the resistivity of a manganin wire at room temperature. 
$^{195}$Pt spectra were deduced by summation of the fast Fourier transform of spin echo signals. The nuclear spin-lattice relaxation rate 1/$T_1T$ of $^{195}$Pt was measured by the saturation-recovery method.
	
	\begin{figure}[htbp]
		\includegraphics[width= 9 cm]{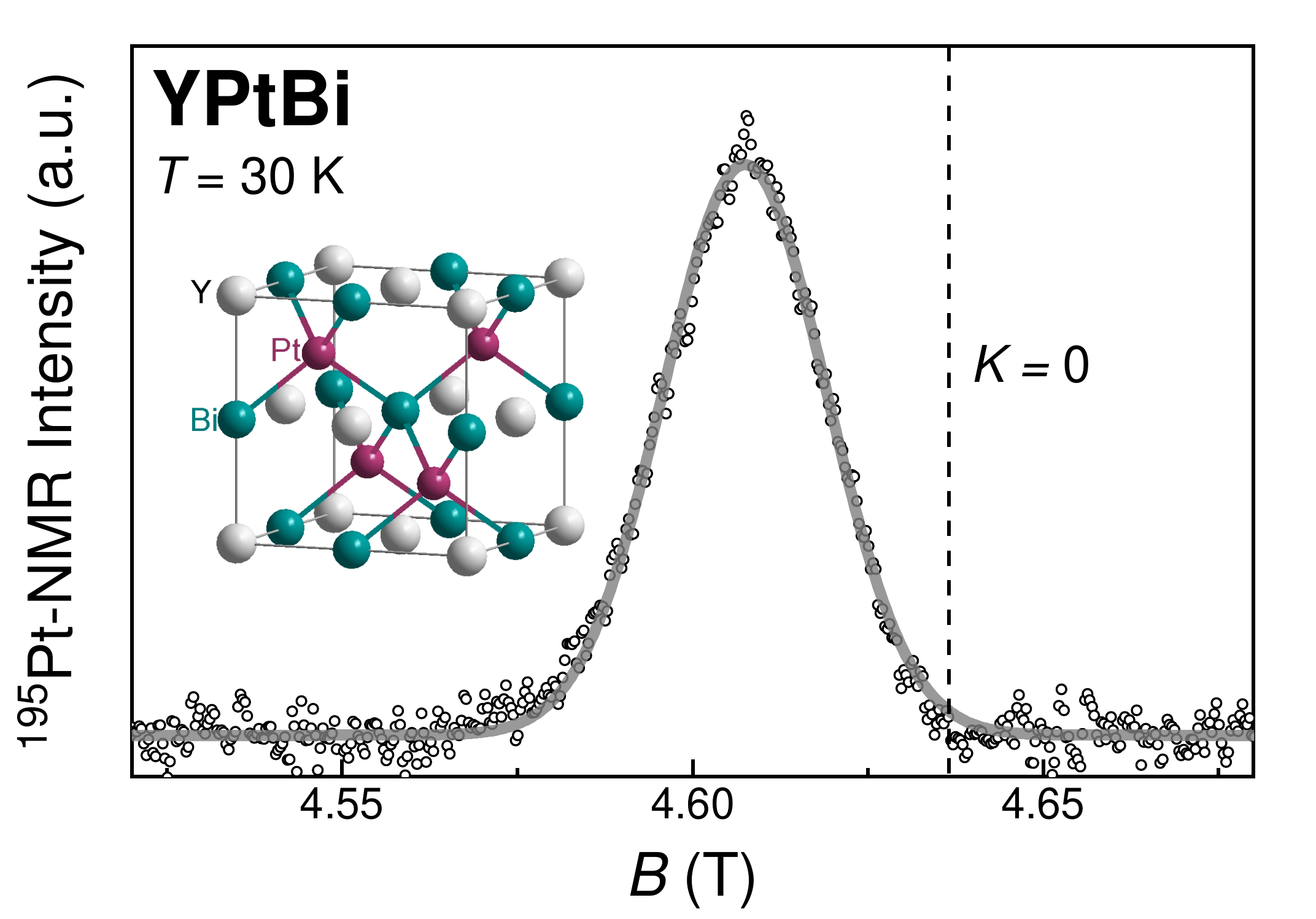}
		\caption{(Color online) The $^{195}$Pt-NMR spectrum obtained by sweeping the magnetic field at the fixed frequency $f_0$ = 42.167 MHz. The grey curve is the fit by a Gaussian function. The dashed line shows Knight shift $K$ = 0. The inset shows the crystal structure of YPtBi.
			\label{spectrum}}
	\end{figure}
	
	Figure 1 shows the $^{195}$Pt-NMR spectrum by sweeping the magnetic field at the fixed frequency $f_{0}$ = 42.167 MHz. Since the crystal structure of YPtBi is cubic and there is only one Pt site\cite{SM}, only one peak is observed in the spectrum.
The Knight shift $K$ is determined by the peak frequency of NMR spectra from $f = \gamma_{n}B(1+K)$, where $^{195}\gamma_{n}$ = 9.094 MHz/T is the gyromagnetic ratio. Figure 2(a) shows the temperature-dependent Knight shifts at various fields. Generally, $K$ is composed of spin and orbital part $K_{\rm s}$ and $K_{\rm orb}$ as $K = K_{\rm s} + K_{\rm orb} = A_{\rm hf}^{\rm spin}\chi_{\rm s} + A_{\rm hf}^{\rm orb}\chi_{\rm orb}$. Here, $A_{\rm hf}^{\rm spin}$ and $A_{\rm hf}^{\rm orb}$ are spin and orbital hyperfine coupling constants, respectively. The $\chi_{\rm s}$ is the  spin susceptibility, and $\chi_{\rm orb}$ is the orbital susceptibility.  $K_{\rm orb}$ is both temperature- and field-independent for materials with only trivial bands. For materials with linearly dispersed bands, the break of the linear response theory will lead to the rise of Landau magnetism, which makes $\chi_{\rm orb}$  field- and temperature- dependence as observed in graphene and TaAs\cite{Nonlinear_magnetization,graphene,Landau_diamagnetism}. In our study, both field- and temperature-dependence of  Knight shift $K$ are barely seen, indicating that the excitations from the topological bands barely contribute to the electronic state at the Femi level at low temperatures. Our results are consistent with the previous ARPES study that the Dirac cone is 300 meV away from the Fermi surface in YPtBi\cite{London,unusual_topological_surface_states}.
	
	\begin{figure}[htbp]
		\includegraphics[width= 9 cm]{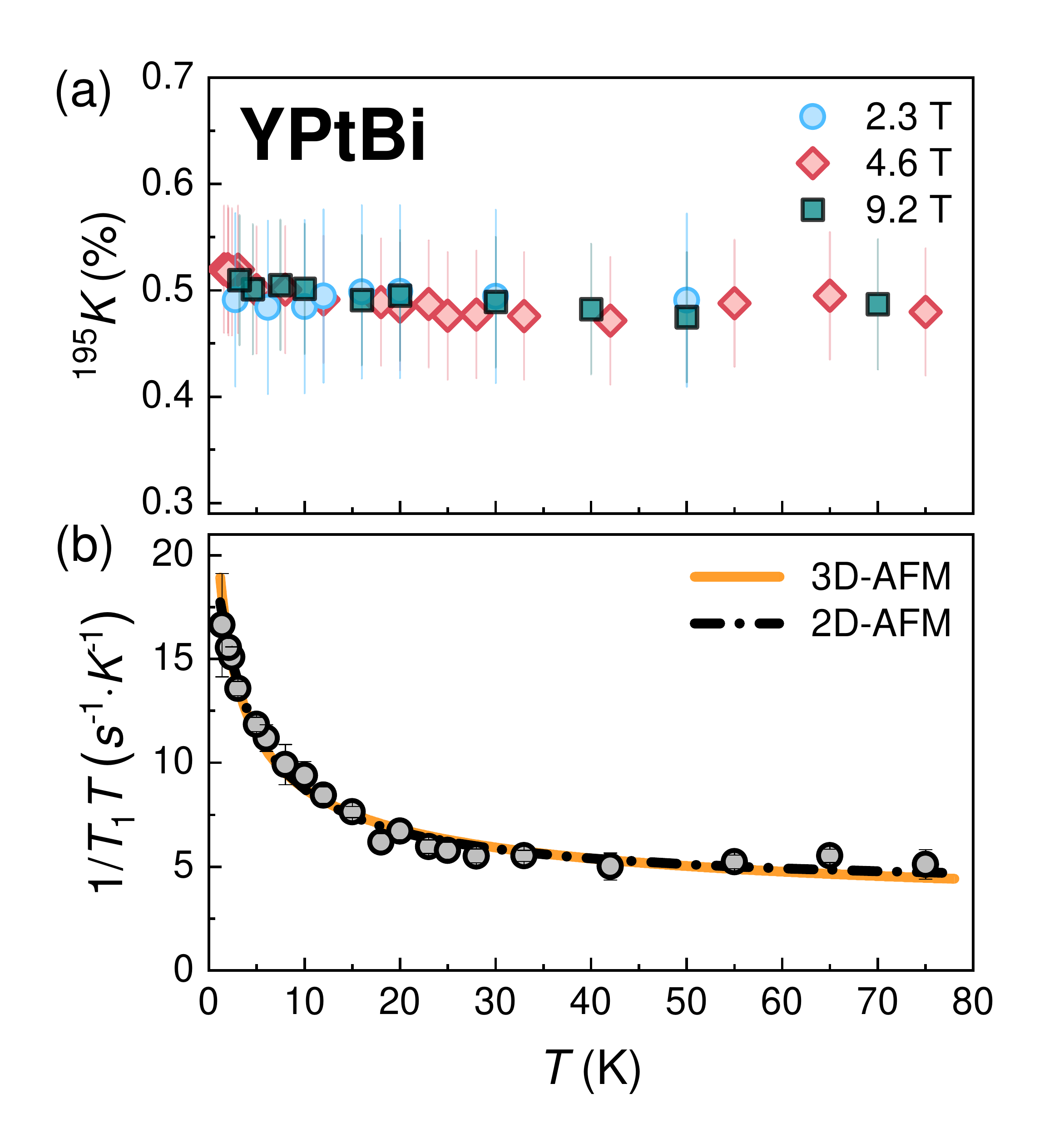}
		\caption{(Color online) (a) Temperature independence of $^{195}K$ at various fields.  (b) Temperature dependent 1/$T_1T$ at 4.6 T. The yellow curve is a fit to $\tfrac{1}{T_1T}={\left (\tfrac{1}{T_1T}\right )_{0}} + {\tfrac{C}{\sqrt{T+\theta}}}$. The dashed black curve is a fit to equation $\tfrac{1}{T_1T}={\left (\tfrac{1}{T_1T}\right )_{0}} + {\tfrac{C}{T+\theta}}$.
			\label{Normal properties}}
	\end{figure}

	Figure 2(b) shows the temperature-dependent 1/$T_1T$ in the normal state. 1/$T_1T$ shows a monotonic increase with decreasing temperature. One possibility is that the electronic band has a branch with a narrow bandwidth close to the Fermi level\cite{Kishimoto1995}. But such band structure was not seen by either the band calculation or ARPES experiments\cite{DOS,unusual_topological_surface_states}. Considering that $^{195}$Pt nuclei do not have a quadrupole moment, the most likely reason for the enhancement of 1/$T_1T$ at low temperatures is due to the existence of spin fluctuations. We note that there is a small increase in Knight shift with decreasing temperature in Fig. \ref{Normal properties}(a). However, the core polarization effect from the $d$-electrons of Pt causes the hyperfine coupling constant $A_{\rm hf}$ to be negative, as observed in other half-Heusler compounds\cite{Nowak2008,Koyama2011}. The increase in Knight shift suggests a decrease in spin susceptibility, indicating antiferromagnetic (AF) spin fluctuations.
	
	The temperature dependence of 1/$T_1T$ can be written as, ${\dfrac{1}{T_1T}} = {\left (\dfrac{1}{T_1T}\right )_{0}} + {\left (\dfrac{1}{T_1T}\right )_{\rm AF}}$. Here, $(1/T_1T)$$_{0}$ represents the intraband contribution, related to the density of states at the Fermi level. $(1/T_1T)_{\rm AF}$ represents the contribution from the AF spin fluctuations. In the case of 2D and 3D AF spin fluctuations, $(1/T_1T)_{\rm AF}$ can be further transcribed as a Curie-Weiss-like formula in different forms\cite{Moriya_2000},
	\begin{equation}
		\label{eq2}
		(1/T_1T)_{\rm AF}\propto\left\{
		\begin{matrix}
			\chi_{q}(T)\propto{\dfrac{C}{T+\theta}}, & \quad \text{(2D AF fluctuations),}\\[5mm]
			\sqrt{\chi_{q}(T)}\propto{\dfrac{C}{\sqrt{T+\theta}}}, & \quad  \text{(3D AF fluctuations)}
		\end{matrix}
		\right.
	\end{equation}
	where $\theta$ is a symbolic temperature related to AF correlations. As shown in Figure 2(b), the experimental data can be well-fitted by these equations of 2D and 3D AF spin fluctuations, so we can not distinguish the dimensionality of the spin fluctuations. In ternary half-Heusler compounds, antiferromagnetic order was observed due to the Ruderman-Kittel-Kasuya-Yosida interaction between conduction electrons and localized moments from the rare earth ions\cite{Nakajima2015}. However, as $Y^{3+}$ ions do not carry any local moments, the observed spin fluctuations might be related to the electron correlations arising from Pt 5$d$ orbitals. The physical origin of such AF spin fluctuations requires further theoretical and experimental investigations.

	\begin{figure}[htbp]
		\includegraphics[width= 9 cm]{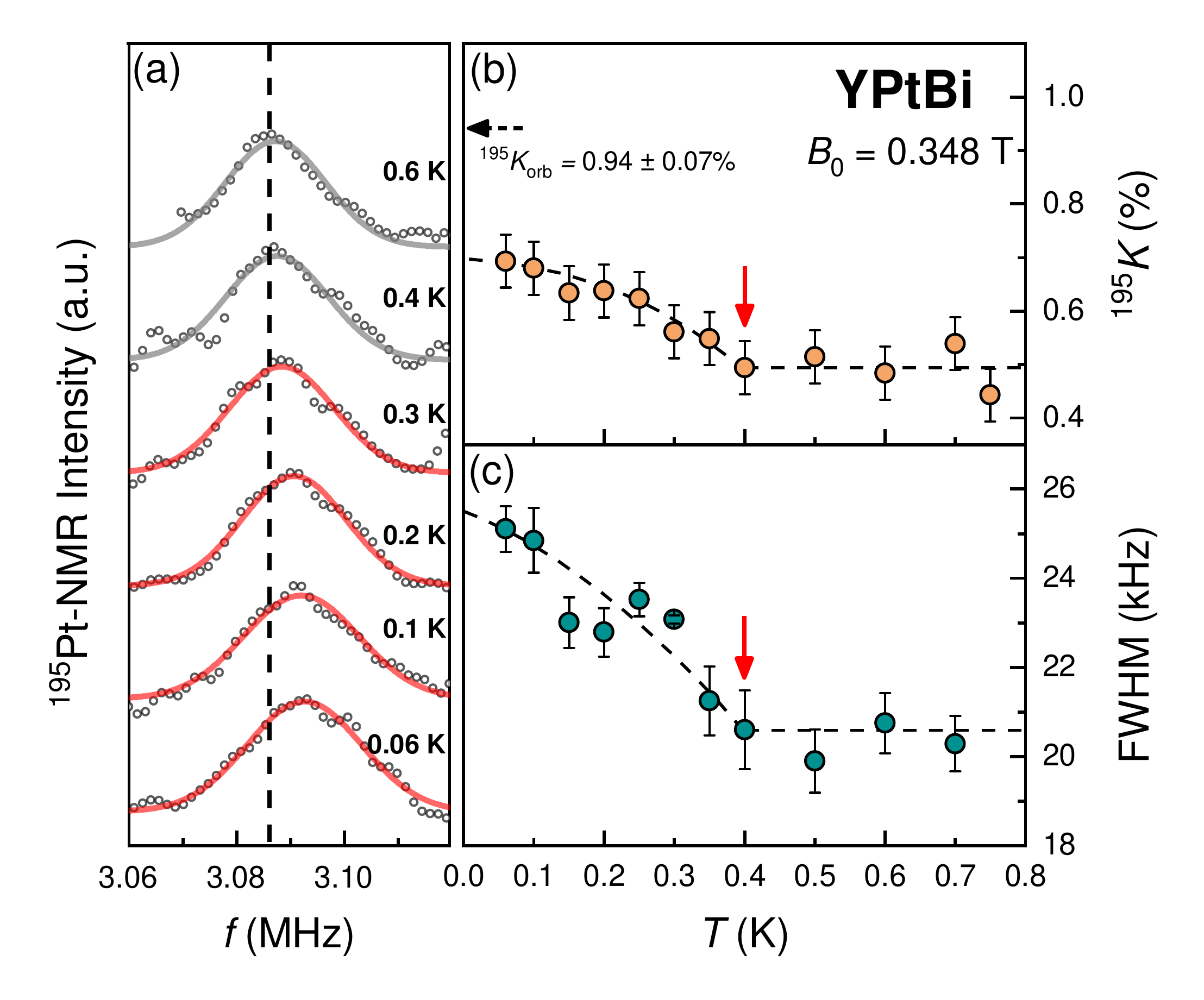}
		\caption{(Color online) (a) The $^{195}$Pt-NMR spectra at various temperatures at 0.348 T. The solid curves are fitted by the Gaussian function. (b) The temperature dependence of the $^{195}$Pt Knight shift $K$ at 0.348 T. $T_{\rm c}$ is marked with a red arrow. The horizontal arrow indicates the position of the orbital Knight shift. Dashed lines are guides for eyes.
	\label{Superconducting spectra}}
	\end{figure}
	
	Next, we turn to the superconducting state of YPtBi.
Figure 3(a) shows the $^{195}$Pt-NMR spectra at various temperatures at $B_0$ = 0.348 T. A line broadening and a shift to the higher frequency below $T_{\rm c}$ is clearly observed. Figure 3(b) shows the Knight shift $K$ as a function of temperature.  We estimated $K_{\rm dia}$ from the superconducting diamagnetic shielding effect\cite{De_Gennes_1999}, which is only about -0.00219$\%$ and can be neglected in the following analysis\cite{SM}. Therefore, considering that $A_{hf}$ is negative, the increase of Knight shift in the superconducting state indicates the decrease of $\chi_{\rm spin}$. Our observation is in contrast to the non-centrosymmetric Li$_{2}$Pt$_{3}$B where the Knight shift does not change below $T_{\rm c}$\cite{Li2Pt3B}, but is similar to Li$_{2}$Pd$_{3}$B which has the same structure as Li$_2$Pt$_3$B but a smaller SOC\cite{Li2Pd3B}.
However, in order to know whether the superconducting pairing is of spin-singlet dominant, one needs to know $K_{\rm orb}$. The common approach is to plot $K$ as a function of $\chi$ to deduce the contribution from $\chi_{\rm spin}$ in the total Knight shift $K$\cite{Matano2008}. But this method is not possible for YPtBi, since Knight shift is nearly $T$-independent in the normal state.

Another method to obtain $K_{\rm orb}$ is to change the density of state (DOS) at the Fermi level, $N(E_F)$, by doping or applying pressure. Since $K_s$ is proportional to $N(E_F)$ and $(1/T_1T)$$_{0}$ is proportional to $N(E_F)^2$, $K_{\rm orb}$ can be deduced by plotting $K$ versus (1/$T_1$$T$)$_{0}$$^{1/2}$\cite{Harada2012}. For YPtBi, we found that both $K$ and 1/$T_1$$T$ have a pressure dependence\cite{SM}. Since 1/$T_1$$T$ is temperature-independent above $T \sim$ 40 K, where $K$ is temperature-independent as well, we have taken $^{195}$$K$ at 40 K and (1/$T_1$$T$)$^{1/2}$ at 65 K  and shown the data for various pressures in Fig. 4. A linear relationship between the two quantities is observed. From the extrapolation of the straight line, we have determined that $^{195}K_{\rm orb} \sim$  $0.94 \pm 0.07 \%$. For a pure spin-singlet state, $K_s$ vanishes at zero temperature. Therefore, our results show that $K_s$ is finite at zero-temperature limit (see Fig. \ref{Superconducting spectra}(b)), suggesting a mixed state of spin-singlet and spin-triplet(or spin-septet) pairing.

\begin{figure}[htbp]
		\includegraphics[width= 9 cm]{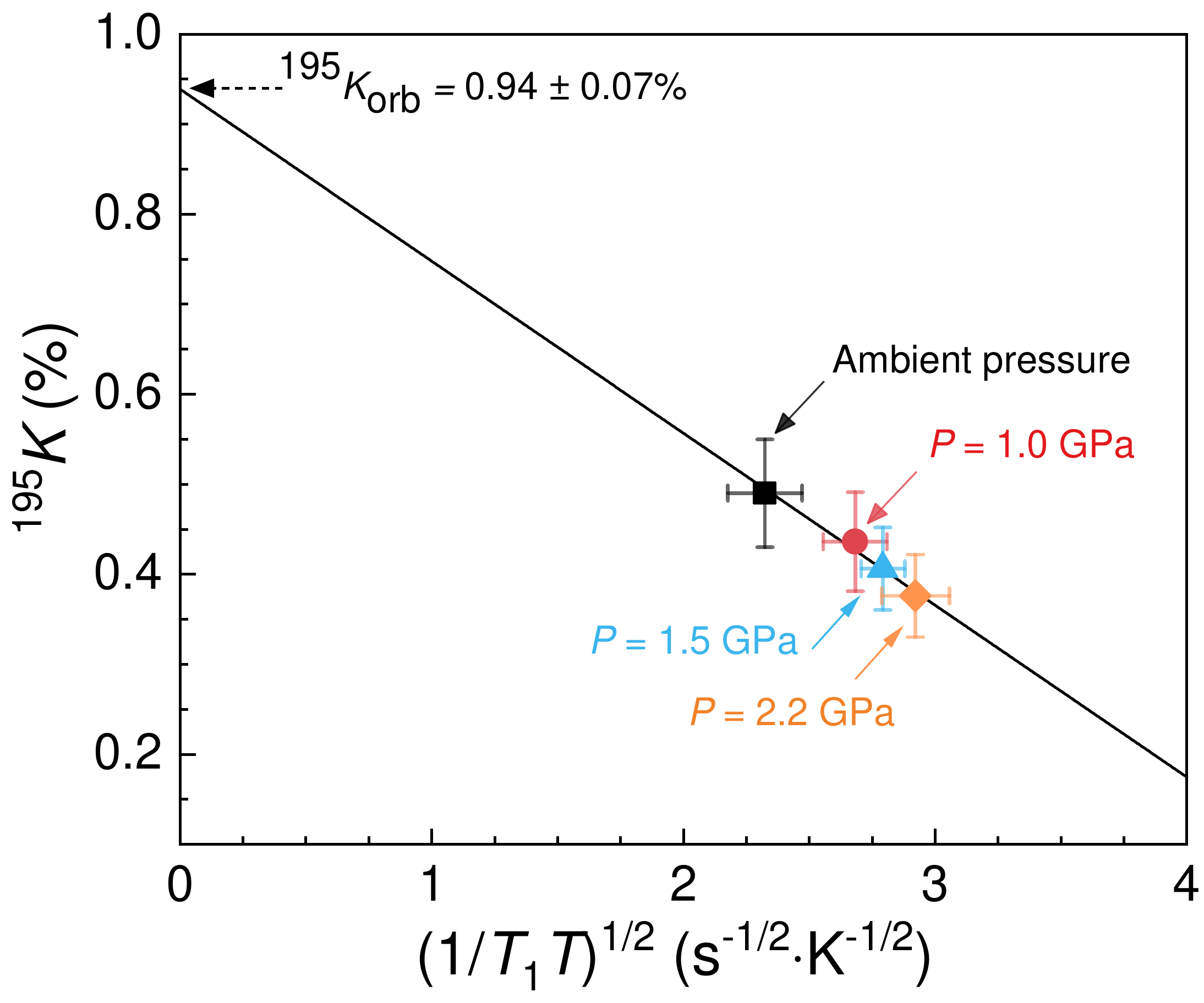}
		\caption{(Color online) The $^{195}$Pt Knight shift in the normal state vs (1/$T_1$$T$)$^{1/2}$ at different pressures, in which $^{195}$$K$ was measured at 40 K and  1/$T_1$$T$ was measured at 65 K.
	\label{Korb}}
	\end{figure}	

	The evolution of the NMR line width below $T_{\rm c}$ also suggests an unconventional superconducting state with a substantial component of spin triplet or spin septet pairing. Figure 3(c) shows the full width at half maximum(FWHM) of $^{195}$Pt-NMR spectra as a function of temperature. The FWHM is temperature-independent above $T_{\rm c}$, and starts to increase just below $T_{\rm c}$. One possibility for the broadening is due to the vortices in the superconducting state\cite{Curro2000,Mitrovi2001}. However, by taking the London penetration depth from the previous $\mu$SR measurement\cite{Low_field_1}, we estimated the broadening as $\Delta$FWHM = $10^{-4}$ kHz\cite{SM}, which is far less than our observation of 5 kHz. Therefore, the observed broadening implies the emergence of a distribution of the Knight shift in the superconducting state. The most possible reason is that the pairing has a substantial component of the odd parity pairing. For the odd-parity paring with the $d$-vector along a certain crystal axis, the Knight shift is unchanged below $T_{\rm c}$ with the magnetic field applied perpendicular to the $d$-vector, but changes to $K_{\rm orb}$ towards zero temperature when the magnetic field is along the $d$-vector\cite{CuxBi2Se3_NMR,Yang2021}. In the powder sample used in our work, the orientation of grains is random with no preferred direction. A distribution of the angle between the applied magnetic field and the $d$-vector will lead to the broadening of the spectrum\cite{SM}.

	\begin{figure}[htbp]
		\includegraphics[width= 9 cm]{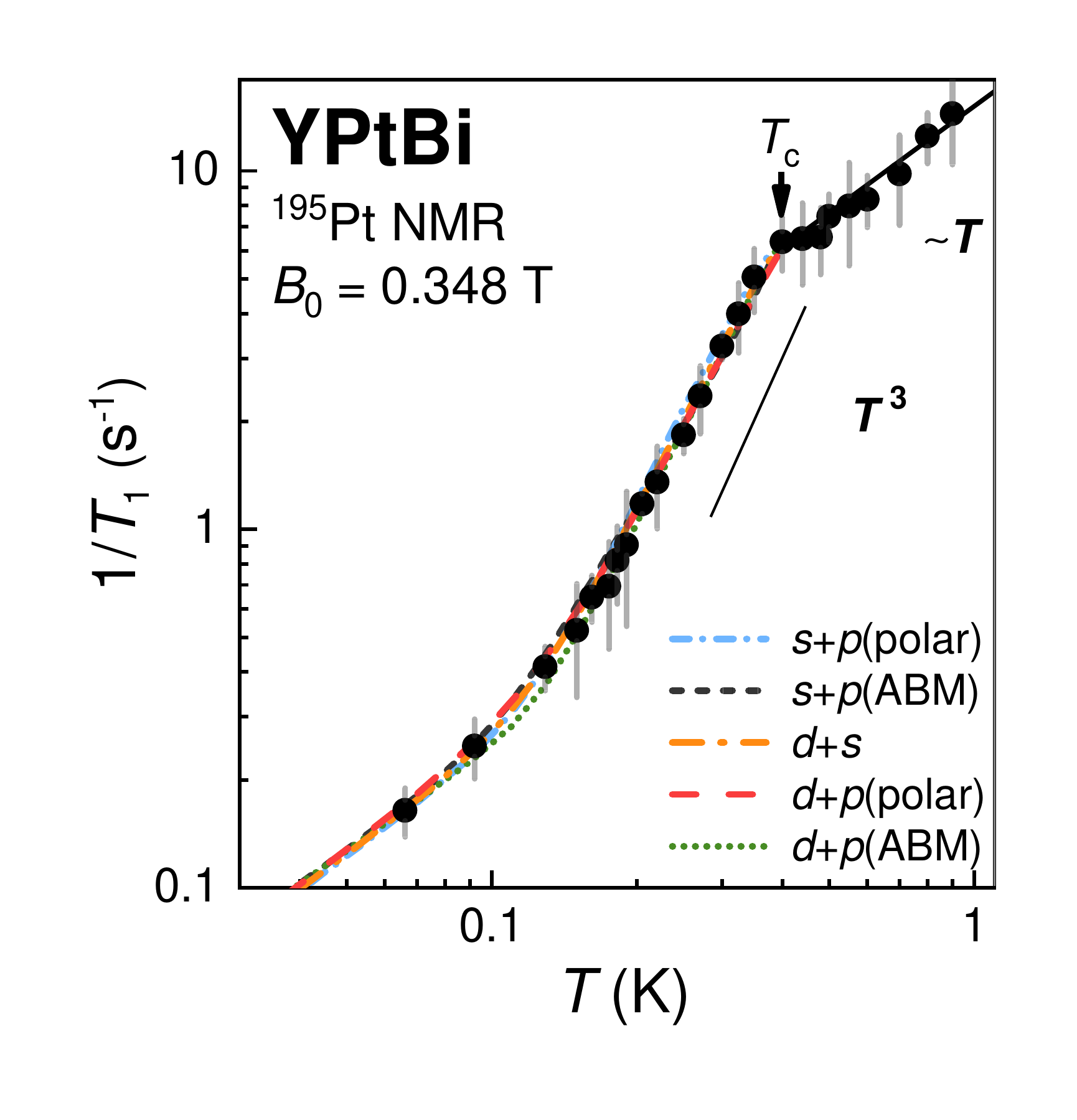}
		\caption{(Color online) 1/$T_1$ as a function of temperature. $T_{\rm c}$ is marked with an arrow. The blue and black dashed curves represent fits with the $s$ + $p$(polar) and $s$ + $p$(ABM) two-gap models(see the text), while the orange dashed curve represents a fit with the $d$ + $s$ two-gap model(see the text). The red dashed and green dotted curves are fits with the $d$ + $p$(polar) and $d$ + $p$  (ABM) two-gap models(see the text), respectively. Thin lines for $T^{3}$ and $T$ are guides for eyes.
			\label{Superconducting 1/T1}}
	\end{figure}
	
	The unconventional nature of the superconductivity also manifests in the temperature dependence of 1/$T_1$, which is shown in Fig. 5. Below $T_{\rm c}$, 1/$T_1$ drops rapidly with the $T^{3}$ dependence, with no Hebel-Slichter coherence peak, suggesting the unconventional nature of the superconductivity. This is in sharp contrast to the isostructural compound LaPtBi where a clear Hebel-Slichter peak was observed\cite{Matano2013}. The (1/$T_{1s}$) in the superconducting state is expressed as\cite{Sigrist1991},
\begin{equation}	
		{\dfrac{T_{1N}}{T_{1S}}} = \dfrac{2}{k_{B}T}\iint	{\left (1 + \dfrac{\Delta^{2}}{EE'}\right )} N_{s}(E)N_{s}(E')f(E)[1-f(E')]\times \delta (E-E')dEdE'
\end{equation}

where 1/$T_{1N}$ (1/$T_{1S}$) is the relaxation rate in the normal (superconducting) state, $N_{s}(E) = N_0\cdot E/\sqrt{E^2-\Delta^2} $ is the superconducting DOS, $f(E)$ is the Fermi distribution function, and $C = 1 + \dfrac{\Delta^{2}}{EE'}$ is the coherence factor.
For an $s$-wave gap, the coherence factor and the divergence of the DOS at $E = \Delta$ will lead to a Hebel-Slichter peak just below $T_{\rm c}$.
If there are nodes in the gap function, the term ${\Delta^{2}}/{EE'}$ disappears due to sign change, then the coherence peak will not be observed, as in cuprates\cite{Lee2006}. The important feature of $1/T_1 \varpropto T^3$ is consistent with the existence of line nodes in the gap function. For example, in the $d$-wave model, 
$N_S(E) \varpropto E$ at low $E$, which results in a $T^3$ variation of 1/$T_1$ following Eq. (2). Meanwhile, we notice that the temperature-dependent 1/$T_1$ shows a crossover from $T^{3}$ to $T$-linear behavior below $T \sim 0.3 T_c$, suggesting the existence of impurity scattering\cite{Bang2004}, which is always observed in the superconductor with line nodes in the gap function such as cuprates and Na$_{0.35}$CoO$_{2}\cdot y$H$_{2}$O\cite{Lee2006,Na0.35CoO2}. However, we note that a simple $d$-wave model is not consistent with the London Penetration depth measurement, which shows that multiple gaps exist in the superconducting state\cite{London}.

So we try to fit the data with multiple gaps, assuming that the total superconducting DOS is contributed from two gaps as $N_{\rm tot} =  \alpha \cdot N_1 + ( 1-\alpha ) \cdot N_2$, where $N_1$ and $N_2$ are the superconducting DOS from two  gaps.
First, we employed the $s$ + $p$ two-gap model\cite{London,singlet_septet,Majorana_surface_modes,Inflated_nodes}. By considering the resonant impurity scattering\cite{Hirschfeld1986}, we tried both ABM and polar type $p$-wave order parameter, as $\Delta^{\rm polar} \left ( \theta, \varphi   \right ) =\Delta_0 {\rm cos}\theta$ and $\Delta^{\rm ABM} \left ( \theta, \varphi   \right ) =\Delta_0 {\rm sin}\theta \cdot e^{i\varphi }$ as shown in Fig. 5\cite{Anderson1961}. The blue dashed curves is a fit by using the $s$ + $p$(polar) two-gap model with $\Delta^1_{s0}=2.0k_{\rm B}T_{\rm c}, \Delta^2_{p0}=2.9k_{\rm B}T_{\rm c}$, $\alpha=0.43$ and $\Gamma/\Delta_{0}=0.12$.
The black dashed curves is a fit by using the $s$ + $p$(ABM) two-gap model with $\Delta^1_{s0}=1.8k_{\rm B}T_{\rm c}, \Delta^2_{p0}=3.2k_{\rm B}T_{\rm c}$, $\alpha=0.45$ and $\Gamma/\Delta_{0}=0.20$. Our calculations show that $1/T_1$ can be fitted very well by using the $s$ + $p$ two-gap model. Next, we tested the $d$ + $s$ two-gap model\cite{Singlet-quintet,Singlet-quintet_2,d+s}. With $\Delta^1_{d0}=1.3k_{\rm B}T_{\rm c}, \Delta^2_{s0}=3.2k_{\rm B}T_{\rm c}$, $\alpha=0.52$ and $\Gamma/\Delta_{0}=0.2$, 1/$T_1$ can still be fitted very well(see Fig. 5). In the end, we utilized  $d$ + $p$ two-gap model, and found that our results can also be well-fitted as demonstrated in Fig. 5. For the $d$ + $p$(polar) two-gap model, we took $\Delta^1_{d0}=2.4k_{\rm B}T_{\rm c}, \Delta^2_{d0}=1.8k_{\rm B}T_{\rm c}$, $\alpha=0.50$ and $\Gamma/\Delta_{0}=0.08$. For the $d$ + $p$(ABM) two-gap model, we took  $\Delta^1_{d0}=2.4k_{\rm B}T_{\rm c}, \Delta^2_{d0}=1.6k_{\rm B}T_{\rm c}$, $\alpha=0.50$ and $\Gamma/\Delta_{0}=0.12$.  We note that electron correlation effects are not considered in our 1/$T_1$ simulation model. Future studies are needed to include electron correlation effects in the simulation model to further improve the understanding of the superconducting properties of YPtBi.

Although our calculations show that various two-gaps models can fit our 1/$T_1$ data very well,  the residual spin susceptibility at zero-temperature and the anomalous line broadening in the superconducting state suggest that a mixed state of $s$ or $d$-wave singlet and $p$-wave triplet(or septet) pairing is more possible. An admixture of $s$-wave singlet and $p$-wave septet superconducting states was proposed in LuPbBi\cite{Ishihara2021}. Theoretical calculation suggests a large band splitting by SOC of 500 meV in LuPdBi, which is similar to 680 meV in YPtBi\cite{zhang2016}. In non-centrosymmetric half-Heusler compound LaPtBi, the band splitting by SOC at the Fermi level is only a few meV\cite{Oguchi2001} and an isotropic superconducting gap is dominant\cite{Matano2013}. All these immediately indicate that SOC is crucial for forming unconventional superconductivity in non-centrosymmetric half-Heusler compounds. We note that $d$ + $p$ two-gaps model was not proposed by any current theory so far. However, the observed AF spin fluctuations could help forming the $d$-wave pairing as observed in cuprates\cite{Lee2006}, which suggests that $d$ + $p$ pairing symmetry is more self-consistent in describing the data for both the superconducting and normal states in YPtBi. In any case, our results show that YPtBi is a unique platform to study the relationship between unconventional superconductivity, topological band structure, and spin fluctuations.

In summary, through the $^{195}$Pt-NMR measurements in YPtBi, we have shown that YPtBi is an unconventional superconductor with AF spin fluctuations in the normal state. The residual spin susceptibility at zero-temperature limit, as well as a large increase of line width in the superconducting state, suggest that it is a mixed state of the spin-singlet with a substantial component of spin-triplet or spin-septet pairing. Moreover, we find that the temperature dependence of 1/$T_1$ can be explained by the $s$ + $p$ or $d$ + $p$ pairing symmetry. Our finding provide new insights into the TSC candidate YPtBi and important constraints for the theoretical modeling of such promising superconducting phase. We hope that these results will stimulate further experimental and theoretical works on the half-Heusler superconductors.

\begin{acknowledgments}
This work was supported by the National Natural Science Foundation of China (Grant Nos. 11974405), the National Key Research and Development Projects of China (Grant Nos. 2022YFA1403400 and Nos. 2022YFA1402600), the Strategic Priority Research Program of the Chinese Academy of Sciences (Grant Nos. XDB33010100) and the Interdisciplinary program of Wuhan National High Magnetic Field Center (Grant No. WHMFC202126), Huazhong University of Science and Technology. Y. F. Guo acknowledges the open project of Beijing National Laboratory for Condensed Matter Physics (Grant No. ZBJ2106110017) and the Double First-Class Initiative Fund of ShanghaiTech University. A portion of this work was carried out at the Synergetic Extreme Condition User Facility (SECUF).
\end{acknowledgments}


\begin{references}
	
	\bibitem{Topologicalinsulators}
	M. Z. Hasan and C. L. Kane,  \href{https://journals.aps.org/rmp/abstract/10.1103/RevModPhys.82.3045}{\emph{Rev. Mod. Phys.} {\bf 82}, 3045 (2010).}
	
	\bibitem{Topologicalinsulatorsandsuperconductors}
	X. L. Qi and S. C. Zhang,  \href{https://journals.aps.org/rmp/abstract/10.1103/RevModPhys.83.1057}{\emph{Rev. Mod. Phys.} {\bf 83}, 1057 (2011).}
	
	\bibitem{PhysRevLett.94.166802}
	S. Das Sarma, M. Freedman, and C. Nayak, \href{https://journals.aps.org/prl/abstract/10.1103/PhysRevLett.94.166802} { \emph{Phys. Rev. Lett.} {\bf 94}, 166802 (2005).}
	
	\bibitem{Sato_2017}
	M. Sato and Y. Ando,  \href{https://iopscience.iop.org/article/10.1088/1361-6633/aa6ac7} { \emph{Rep. Prog. Phys.} {\bf 80}, 076501 (2017).}
	
	\bibitem{CuxBi2Se3_NMR}
	K. Matano, M. Kriener, K. Segawa, Y. Ando, and G.-q. Zheng,  \href{https://www.nature.com/articles/nphys3781} {\emph{Nat. Phys} {\bf 12}, 852 (2016).}
	
	\bibitem{Ran2019}
	L. Jiao, S. Howard, S. Ran, Z. Wang, J. O. Rodriguez, M. Sigrist, Z. Wang, N. P. Butch and V. Madhavan,  \href{https://doi.org/10.1038/s41586-020-2122-2} {\emph{Nature} {\bf 579}, , 523 (2020).}



\bibitem{Yang2021}
	J. Yang,  J. Luo, C. J. Yi, Y. G. Shi, Y. Zhou, And G.-q. Zheng,  \href{https://doi.org/10.1126/sciadv.abl4432} {\emph{Sci. Adv.} {\bf 7}, eabl4432 (2021).}


\bibitem{Li2Pt3B}
	M. Nishiyama, Y. Inada, and G.-q. Zheng, \href{https://journals.aps.org/prl/abstract/10.1103/PhysRevLett.98.047002}{\emph{Phys. Rev. Lett.} \textbf{98}, 047002 (2007).}


\bibitem{PhysRevB.76.045302}
	L. Fu and C. L. Kane,  \href{https://journals.aps.org/prb/abstract/10.1103/PhysRevB.76.045302} {\emph{Phys. Rev. B}  \textbf{76}, 045302 (2007).}

	
	\bibitem{HgTequantumwells}
	B. A. Bernevig, T. L. Hughes, and S. C. Zhang, \href{https://www.science.org/doi/10.1126/science.1133734}{\emph{Science} {\bf 314}, 1757 (2006).}
	
\bibitem{Half-Heusler}
	H. Lin, L. A. Wray, Y. Xia, S. Xu, S. Jia, R. J. Cava, A. Bansil, and M. Z. Hasan, \href{https://www.nature.com/articles/nmat2771} {\emph{Nat. Mater.}  \textbf{9}, 546 (2010).}
	
	\bibitem{Tunable}
	S. Chadov, X. L. Qi, J. Kubler, G. H. Fecher, C. Felser, and S. C. Zhang, \href{https://www.nature.com/articles/nmat2770} {\emph{Nat. Mater.} \textbf{9}, 541 (2010)}.
	


	
	
	\bibitem{PhysRevLett.105.096404}
	D. Xiao, Y. G. Yao, W. X. Feng, J. Wen, W. G. Zhu, X. Q. Chen, G. M. Stocks, and Z. Y. Zhang, \href{https://journals.aps.org/prl/abstract/10.1103/PhysRevLett.105.096404} {\emph{Phys. Rev. Lett.} \textbf{105}, 096404 (2010).}
	

	\bibitem{PhysRevB.82.125208}
	W. Al-Sawai, H. Lin, R. S. Markiewicz, L. A. Wray, Y. Xia, S.-Y. Xu, M. Z. Hasan, and A. Bansil, \href{https://journals.aps.org/prb/abstract/10.1103/PhysRevB.82.125208} {\emph{Phys. Rev. B} {\bf 82}, 125208 (2010).}
	
	\bibitem{unusual_topological_surface_states}
	Z. K. Liu, L. X. Yang, S.-C. Wu, C. Shekhar, J. Jiang, H. F. Yang, Y. Zhang, S.-K. Mo, Z. Hussain, B. Yan, C. Felser, and Y. L. Chen, \href{https://www.nature.com/articles/ncomms12924}{\emph{Nat. Commun.} {\bf 7}, 12924 (2016)}.
	
	\bibitem{Metallic_surface_electronic_state}
	C. Liu, Y. Lee, T. Kondo, E. D. Mun, M. Caudle, B. N. Harmon, S. L. Bud'ko, P. C. Canfield, and A. Kaminski,  \href{https://journals.aps.org/prb/abstract/10.1103/PhysRevB.83.205133}{\emph{Phys. Rev. B} {\bf 83}, 205133 (2011)}.
	
	\bibitem{Superconductivity}
	N. P. Butch, P. Syers, K. Kirshenbaum, A. P. Hope, and J. Paglione,  \href{https://journals.aps.org/prb/abstract/10.1103/PhysRevB.84.220504} {\emph{Phys. Rev. B} {\bf 84}, 220504(R) (2011).}
	
	\bibitem{LaBiPt_SC}
	G. Goll, M. Marz, A. Hamann, T. Tomanic, K. Grube, T. Yoshino, and T. Takabatake,   \href{https://doi.org/10.1016/j.physb.2007.10.089} {\emph{Phys. B: Condens. Matter} {\bf 403}, 1065 (2008)}.
		
	\bibitem{LnBiPt_SC}
	F. F. Tafti, T. Fujii, A. Juneau-Fecteau, S. Ren\'{e} de Cotret, N. Doiron-Leyraud, A. Asamitsu, and L. Taillefer,  \href{https://journals.aps.org/prb/abstract/10.1103/PhysRevB.87.184504}{\emph{Phys. Rev. B} {\bf 87}, 184504 (2013)}.
	
	\bibitem{pressure}
	T. V. Bay, T. Naka, Y. K. Huang, and A. de Visser,   \href{https://journals.aps.org/prb/abstract/10.1103/PhysRevB.86.064515}{\emph{Phys. Rev. B} {\bf 86}, 064515 (2012)}.
	

	\bibitem{DOS}
	M. Meinert,  \href{https://doi.org/10.1103/PhysRevLett.116.137001}{\emph{Phys. Rev. Lett.} {\bf 116}, 137001 (2016)}.

	\bibitem{PhysRevLett.117.075301}
	W. Yang, Y. Li, and C. Wu,  \href{https://journals.aps.org/prl/abstract/10.1103/PhysRevLett.117.075301}{\emph{Phys. Rev. Lett.} {\bf 117}, 075301 (2016).}
	
	\bibitem{PhysRevB.96.214514}
	L. Savary, J. Ruhman, J. W. F. Venderbos, L. Fu, and P. A. Lee,  \href{https://journals.aps.org/prb/abstract/10.1103/PhysRevB.96.214514}{\emph{Phys. Rev. B} {\bf 96}, 214514 (2017)}.
	
	\bibitem{Roy_2019}
	B. Roy, S. Ali Akbar Ghorashi, M. S. Foster, and A. H. Nevidomskyy,  \href{https://journals.aps.org/prb/abstract/10.1103/PhysRevB.99.054505}{\emph{Phys. Rev. B} {\bf 99}, 054505 (2019)}.
	

	
	\bibitem{Venderbos_2018}
	J. W. F. Venderbos, L. Savary, J. Ruhman, P. A. Lee, and L. Fu,  \href{https://journals.aps.org/prx/abstract/10.1103/PhysRevX.8.011029} {\emph{Phys. Rev. X} {\bf 8}, 011029 (2018).}

	\bibitem{London}
	H. Kim, K. Wang, Y. Nakajima, R. Hu, S. Ziemak, P. Syers, L. Wang, H. Hodovanets, J. D. Denlinger, P. M. R. Brydon, D. F. Agterberg, M. A. Tanatar, R. Prozorov, and J. Paglione,  \href{https://www.science.org/doi/10.1126/sciadv.aao4513}{\emph{Sci. Adv} {\bf 4}, eaao4513 (2018)}.
	
	\bibitem{singlet_septet}
	P. M. R. Brydon, L. Wang, M. Weinert, and D. F. Agterberg \href{https://journals.aps.org/prl/abstract/10.1103/PhysRevLett.116.177001}{\emph{Phys. Rev. Lett.} {\bf 116}, 177001 (2016)}.
	

	

	\bibitem{Majorana_surface_modes}
	W. Yang, T. Xiang, and C. Wu,  \href{https://journals.aps.org/prb/abstract/10.1103/PhysRevB.96.144514} {\emph{Phys. Rev. B} {\bf 96}, 144514 (2017)}.
	
	\bibitem{Inflated_nodes}
	C. Timm, A. P. Schnyder, D. F. Agterberg, and P. M. R. Brydon,  \href{https://doi.org/10.1103/PhysRevB.96.094526} {\emph{Phys. Rev. B} \textbf{96}, 094526 (2017).}
	
	\bibitem{Singlet-quintet}
	J. Yu and C.-X. Liu,  \href{https://doi.org/10.1103/PhysRevB.98.104514}{\emph{Phys. Rev. B} {\bf 98}, 104514 (2018)}.

	\bibitem{Singlet-quintet_2}
	Q.-Z. Wang, J. Yu, and C.-X. Liu,  \href{https://journals.aps.org/prb/abstract/10.1103/PhysRevB.97.224507}{\emph{Phys. Rev. B} {\bf 97}, 224507 (2018)}.
	
	\bibitem{Boettcher_2018}
	I. Boettcher and I. F. Herbut,  \href{https://journals.aps.org/prl/abstract/10.1103/PhysRevLett.120.057002} {\emph{Phys. Rev. Lett.} {\bf 120}, 057002 (2018).}
	
	\bibitem{d+s}
	G. B. Sim, A. Mishra, M. J. Park, Y. B. Kim, G. Y. Cho, and	S. B. Lee,  \href{https://journals.aps.org/prb/abstract/10.1103/PhysRevB.100.064509} {\emph{Phys. Rev. B} {\bf 100}, 064509 (2019).}

	
	\bibitem{Heat_apacity}
	O. Pavlosiuk, D. Kaczorowski, and P. Wi\'{s}niewski,  \href{https://journals.aps.org/prb/abstract/10.1103/PhysRevB.94.035130} {\emph{Phys. Rev. B} {\bf 94}, 035130 (2016).}
	
      \bibitem{Nowak2014}
	B. Nowak and D. Kaczorowski,  \href{https://journals.aps.org/prb/abstract/10.1103/PhysRevB.94.035130} {\emph{J. Phys. Chem. C} {\bf 118}, 18021 (2014).}

     \bibitem{zhang2016}
     X. Zhang, Z. Hou, Y. Wang, G. Xu, C. Shi, E. K. Liu, X. K. Xi, W. H. Wang, G. H. Wu, and X. X. Zhang, \href{https://www.nature.com/articles/srep23172} {\emph{Sci. Rep.} {\bf 6}, 23172 (2016).}

     \bibitem{SM}
     See Supplemental Material for additional data and analysis.

\bibitem{Yokogawa2007}
K. Yokogawa, K. Murata, , H. Yoshino, and S. Aoyama, Solidification of High-Pressure Medium Daphne 7373. \href{https://doi.org/10.1143/JJAP.46.3636} {\emph{Jpn. J. Appl. Phys.} \textbf{46}, 3636 (2007).}

	\bibitem{Nonlinear_magnetization}
	S. Slizovskiy and J. J. Betouras,  \href{https://journals.aps.org/prb/abstract/10.1103/PhysRevB.86.125440} {\emph{Phys. Rev. B} {\bf 86}, 125440 (2012).}
		
	\bibitem{graphene}
	Z. Li, L. Chen, S. Meng, L. Guo, J. Huang, Y. Liu, W. Wang, and X. Chen, \href{https://journals.aps.org/prb/abstract/10.1103/PhysRevB.91.094429} {\emph{Phys. Rev. B} \textbf{91}, 094429 (2015).}
	
	\bibitem{Landau_diamagnetism}
	C. G. Wang, Y. Honjo, L. X. Zhao, G. F. Chen, K. Matano, R. Zhou, and G.-q. Zheng,  \href{https://journals.aps.org/prb/abstract/10.1103/PhysRevB.101.241110} {\emph{Phys. Rev. B} \textbf{101}, 241110(R) (2020).}
	
    \bibitem{Kishimoto1995}
Y. Kishimoto, T. Ohno, and T. Kanashiro, \href{https://doi.org/10.1143/JPSJ.64.1275} {\emph{J. Phys. Soc. Jpn.} \textbf{64}, 1275 (1995).}
	


    \bibitem{Koyama2011}
T. Koyama, M. Abe, T. Mito, K. Ueda, T. Kohara, and H. S. Suzuki, \href{https://doi.org/10.1143/JPSJS.80SA.SA097} {\emph{J. Phys. Soc. Jpn.} \textbf{80}, SA097 (2011).}

    \bibitem{Nowak2008}
A. Grykalowska, and B. Nowak, \href{https://doi.org/10.1016/j.jallcom.2006.11.072} {\emph{Journal of Alloys and Compounds} \textbf{7}, 453 (2008).}



	\bibitem{Moriya_2000}
	T. Moriya and K. Ueda,  \href{https://www.tandfonline.com/doi/citedby/10.1080/000187300412248?scroll=top&needAccess=true} {\emph{Adv. Phys.} \textbf{49}, 555 (2000).}
	
    \bibitem{Nakajima2015}
Y. Nakajima, R. Hu, K. Kirshenbaum, A. Hughes, P. Syers, X. Wang, K. Wang, R. Wang, S. R. Saha, D. Pratt, J. W. Lynn, and J. Paglione, \href{https://www.science.org/doi/10.1126/sciadv.1500242} {\emph{Sci. Adv.} \textbf{5}, 1500242 (2015).}
	



	

\bibitem{De_Gennes_1999}
P. G. De Gennes \textit{Superconductivity of Metals and Alloys}, (CRC Press, 1999).


\bibitem{Li2Pd3B}
M. Nishiyama, Y. Inada, and G.-q. Zheng,  \href{https://journals.aps.org/prb/abstract/10.1103/PhysRevB.71.220505} {\emph{Phys. Rev. B} \textbf{71}, 220505(R) (2005).}


\bibitem{Matano2008}
K. Matano, C. T. Lin and Guo-qing Zheng, \href{https://doi.org/10.1209/0295-5075/84/57010}{\emph{EPL} \textbf{84}, 57010 (2008).}

\bibitem{Harada2012}
S. Harada, J. J. Zhou, Y. G. Yao, Y. Inada, and Guo-qing Zheng, \href{https://doi.org/10.1103/PhysRevB.86.220502} {\emph{Phys. Rev. B} \textbf{86}, 220502(R) (2012).}



\bibitem{Curro2000}
N.J.Curro, C.Milling, J.Haase, and C. P. Slichter, \href{https://doi.org/10.1103/PhysRevB.62.3473} {\emph{Phys. Rev.B} \textbf{62}, 3473 (2000).}

\bibitem{Mitrovi2001}
V. F. Mitrovi\'{c}, E. E. Sigmund, M. Eschrig, H. N. Bachman, W. P. Halperin, A. P. Reyes, P. Kuhns, and W. G. Moulton, \href{https://doi.org/10.1038/35097039} {\emph{Nature} \textbf{413}, 501 (2002).}

\bibitem{Low_field_1}
T. V. Bay, M. Jackson, C. Paulsen, C. Baines, A. Amato, T. Orvis, M. C. Aronson, Y. K. Huang, and A. de Visser,   \href{https://doi.org/10.1016/j.ssc.2013.12.010} {\emph{Solid State Commun.} \textbf{183}, 13 (2014).}
	



\bibitem{Matano2013}
K. Matano, S. Maeda, H. Sawaoka, Y. Muro, T. Takabatake, B. Joshi, S. Ramakrishnan, K. Kawashima, J. Akimitsu, and G.-q. Zheng, \href{https://doi.org/10.7566/JPSJ.82.084711} {\emph{J. Phys. Soc. Jpn.} \textbf{82}, 084711 (2013).}

\bibitem{Sigrist1991}
M. Sigrist and K. Ueda,  \href{https://doi.org/10.1103/RevModPhys.63.239} {\emph{Rev. Mod. Phys.} \textbf{63}, 239 (1991).}



\bibitem{Lee2006}
K. Asayama, G.-Q. Zheng, Y. Kitaoka, K. Ishida and K. Fujiwara, \href{https://doi.org/10.1103/RevModPhys.78.17} {\emph{Physica C} \textbf{178}, 281 (1991).}

\bibitem{Bang2004}
	Y. Bang, M. J. Graf, A. V. Balatsky, and J. D. Thompson, \href{https://doi.org/10.1103/PhysRevB.69.014505} {\emph{Phys. Rev. B} {\bf 69}, 014505 (2004).}


\bibitem{Na0.35CoO2}
T. Fujimoto, G.-q. Zheng, Y. Kitaoka, R. L. Meng, J. Cmaidalka, and C. W. Chu,  \href{https://journals.aps.org/prl/abstract/10.1103/PhysRevLett.92.047004} {\emph{Phys. Rev. Lett.} {\bf 92}, 047004 (2004).}

\bibitem{Hirschfeld1986}
P. Hirschfeld, D.Vollhardt, and P.W\"{o}lfle, \href{https://doi.org/10.1016/0038-1098(86)90190-0} {\emph{Solid State Commun.} {\bf 59}, 111 (1986).}

\bibitem{Anderson1961}
P. W. Anderson and P. Morel, \href{https://doi.org/10.1103/PhysRev.123.1911}{\emph{Phys. Rev.} \textbf{123}, 1911 (1961).}




\bibitem{Ishihara2021}
K. Ishihara, T. Takenaka, Y. Miao, Y. Mizukami, K. Hashimoto, M. Yamashita, M. Konczykowski, R. Masuki, M. Hirayama, T. Nomoto, R. Arita, O. Pavlosiuk, P. Wi\'{s}niewski, D. Kaczorowski, and T. Shibauchi,  \href{https://journals.aps.org/prx/abstract/10.1103/PhysRevX.11.041048} {\emph{Phys. Rev. X} {\bf 11}, 041048 (2021).}
	

\bibitem{Oguchi2001}
	T. Oguchi, \href{https://doi.org/10.1103/PhysRevB.63.125115} {\emph{Phys. Rev. B} {\bf 63}, 125115  (2001).}


\end{references}
\end{document}